# Visible Spectrum Circular Dichroism in Extrinsic Chirality Metamaterials

*Seoungjun Lee[1], Zengbo Wang[1,2], Cheng Feng[3], Jiao Jiao[1], Ashfaq Khan[4], Lin Li[1]

[1]School of Mechanical, Aerospace and Civil Engineering, University of Manchester, Manchester, M13 9PL, UK

[2]Schools of Electronic Engineering, Bangor University, Dean Street, Bangor, Gwynedd, LL57 1UT, UK

[3]Chemistry Department, University of Leicester, University Road, Leicester, LE1 7RH, UK

[4]Department of Mechanical Engineering, NWFP University of Engineering and Technology, Peshawar, 25000, Pakistan

ABSTRACT: We present the new planar extrinsic chirality metamaterial (ECM) design that manifests giant circular dichroism (CD) in the visible spectrum range rather than usual near-infrared and terahertz range. Effects of incident beam angles and meta-molecules unit sizes on the CD spectrums were theoretically analyzed; Physical mechanism was illustrated in new figures of asymmetrical current excitation in neighboring unit cells.

Circular dichroism (CD) is characterized by the differential absorption of right circularly polarized (RCP) light and left circularly polarized (LCP) light, and is related to optical activity *[1]*. Circular dichroism spectroscopy has been widely used to gain information about biomolecule, DNA *[2-4]* and organic compounds *[5]*. Nano-particles obtained strong CD in the geometry and composition of a chiral molecule *[6-8]*. CD was observed in double layers *[9]* and mutually twisted unconnected layers *[10]* of chiral planar metamaterials at near-infrared wavelengths. Using terahertz frequency, CD was obtained by optical activity and coupling effects *[11, 12]*. A three-dimensional chiral metamaterial with mutually twisted planar metal patterns in parallel planes generated giant CD due to negative index of refraction *[13]*. This paper presents that new planar chiral metamaterial design can achieve strong CD and different energy distribution in unit cells at visible spectrums. The unit cell of the new planar chiral metamaterial is consisted of a silver single layer structure on borosilicate glass substrate.

The transmission of RCP and LCP can be mathematically defined in a 2×2 t-matrix as *[1]*:

$$\begin{pmatrix} T_+ \\ T_- \end{pmatrix} = \begin{pmatrix} t_{++} & t_{-+} \\ t_{+-} & t_{--} \end{pmatrix} \begin{pmatrix} I_+ \\ I_- \end{pmatrix} \quad (1)$$

Where '+' and '-' are symbols for RCP and LCP incident beam, receptively. The matrix is combined by transmission ($T$) and incident ($I$) beam. $t_{++}$ and $t_{--}$ correspond to the transmission amplitude of RCP ($I_+$) and LCP ($I_-$) incident beam. $t_{+-}$ and $t_{-+}$ are calculated by the conversion between RCP and LCP. The incident electric field ($E^{in}$) and transmission electric field ($E^{out}$) are related to the t-matrix and are defined as *[1]*:

$$E_i^{out} = t_{ij} E_j^{in} \quad (2)$$

, and CD is mathematically defined as:

$$CD = |t_{++}|^2 - |t_{--}|^2 \quad (3)$$

CD spectrum of a symmetric non-chiral three dimensional nano structure was explored under oblique incident beams to use modern finite integrate technique (FIT) in frequency domain.

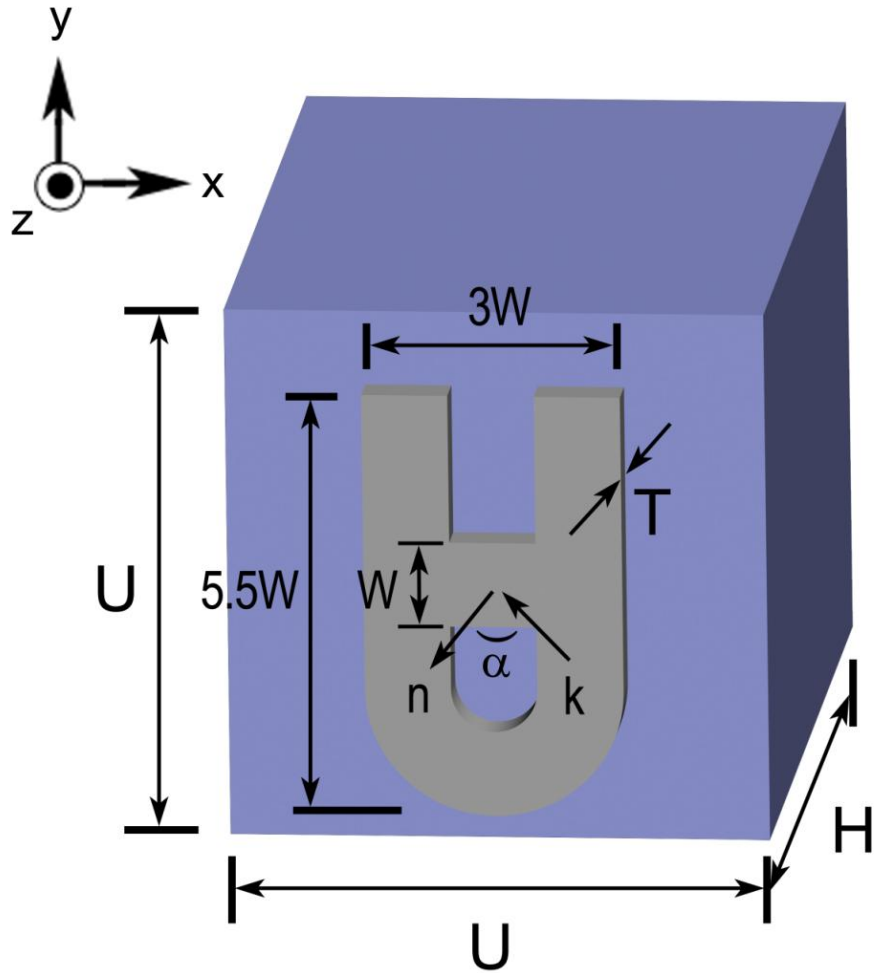

Fig. 1 (Color online) Symmetric 3D unit cell structure (U=250nm) of borosilicate glass substrate (H=1000 nm) and silver unit cells (T=20 nm and W=40nm). *n* is the direction of vector perpendicular to the unit cell, k is the direction of the incident beam and α is angle of the incident beam.

The ECM design consists of unit cells of 20 nm thick silver layer on a 1 μm thick borosilicate glass substrate as shown in Fig. 1. A silver metamaterial structure with borosilicate glass substrate forms X and Y dimension into unit cell structure and Z dimension into open boundary. Figure 2 shows the simulated CD spectrum of the 250 nm unit cell in wavelength between 400 nm and 1200 nm at various

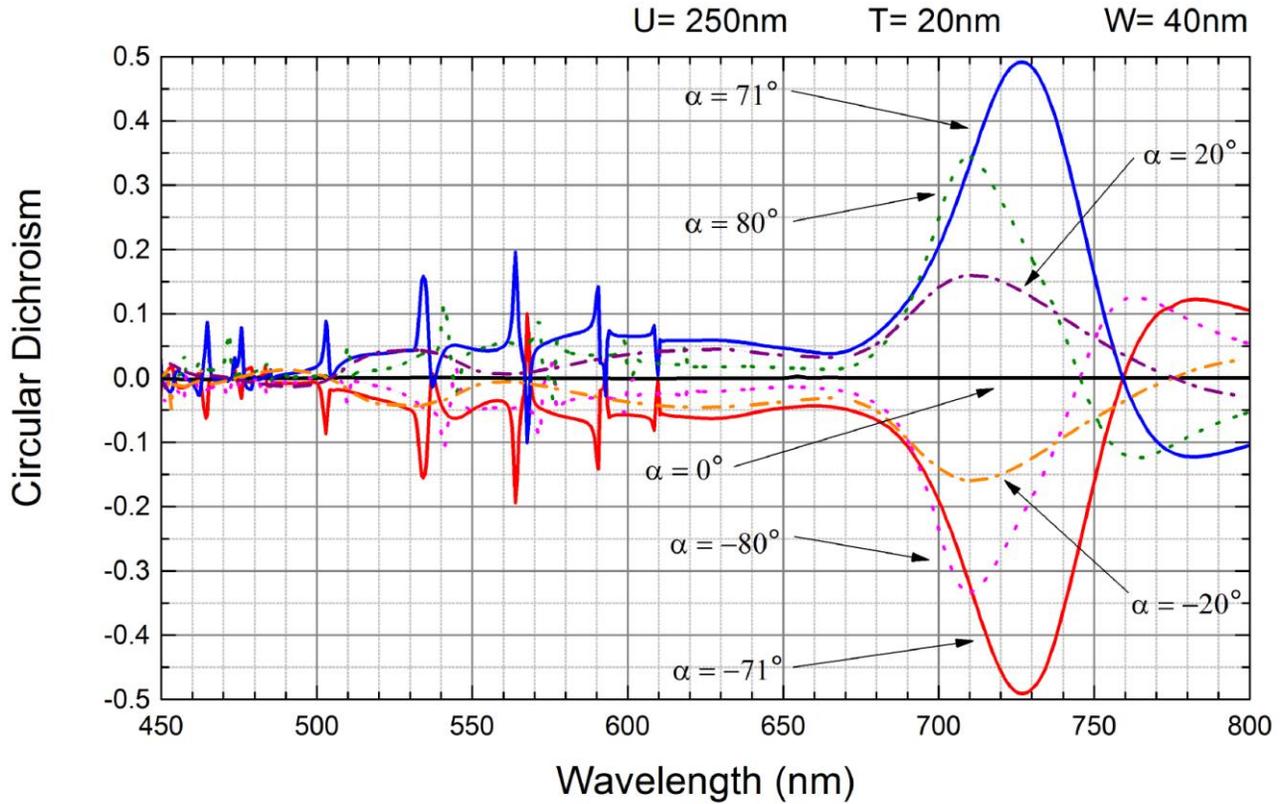

Fig. 2 (Color online) Comparison of CD in the 250 nm unit cell between different incident beam angles as α=±80°, ±71°, ±20° and 0° at wavelength (λ) between 450 nm and 800 nm. The maximum and minimum CDs are 0.4914 and -0.4905 at α=71° and α=-71° with λ=726.39 nm, 0.3438 and -0.3327 at α=80° and α=-80° with λ=710.06 nm and, 0.1602 and -0.1598 at α=20° and α=-20° with λ=710.9 nm, respectively.

incident angles of α=0°, ±20°, ±71° and ±80°. In wavelengths between 400 nm and 600 nm, the spectrum curves are highly oscillating. This phenomenon is due to the strong optical diffraction effect associated with the unit cell *[1]*. Above 600 nm wavelengths, the spectrum curves changes smoothly because the optical diffraction is negligible. The maximum CD peaks was 0.4914 at 726.39 nm visible wavelength, occurring at angles α= ±71°. It shows that ±71° tiled incident beams may have most significant interactions between electric and magnetic dipoles, which may results in generation of most

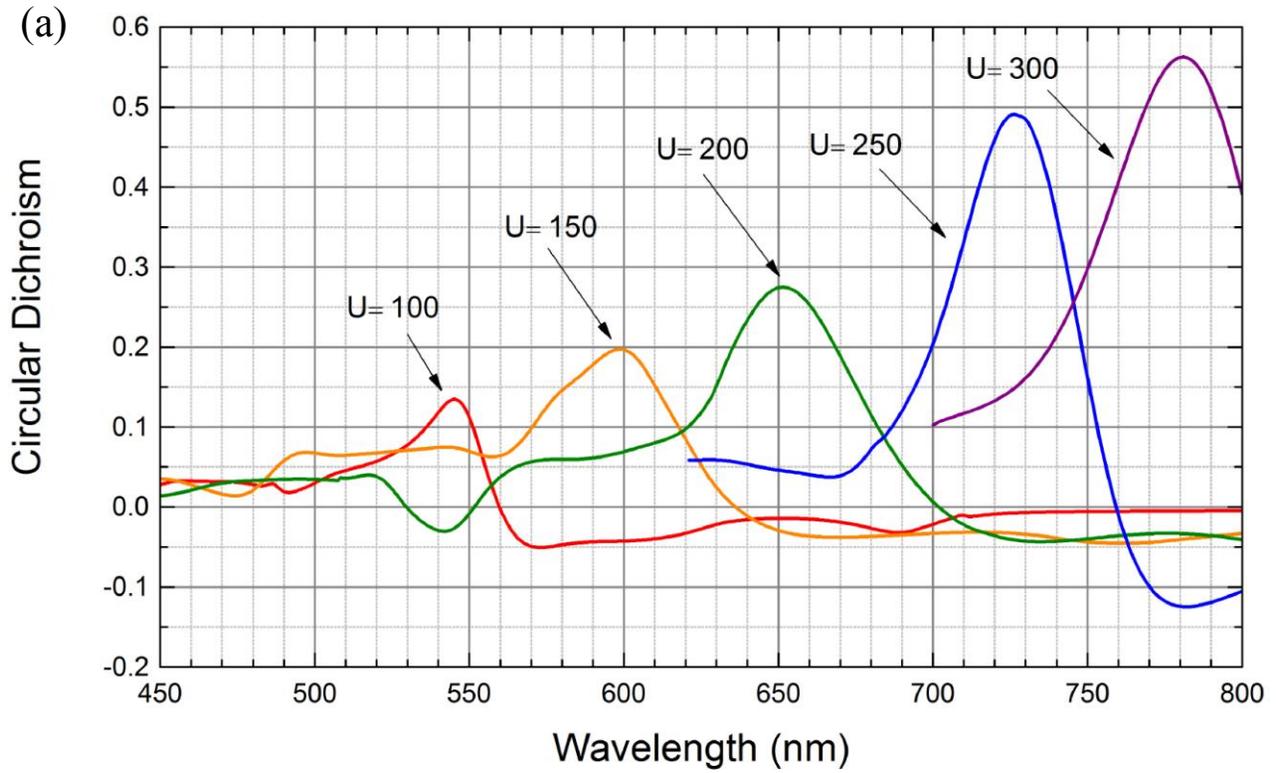

(a)

(b)

| Unit cell size (U, nm) | Optimized angle (α, °) | Optimized wavelength (λ, nm) | Circular dichroism (CD) | |
|---|---|---|---|---|
| 100 | ±77 | 545.45 | +77 | 0.1345 |
| 150 | ±58 | 598.80 | +58 | 0.1976 |
| 200 | ±45 | 651.47 | +45 | 0.2750 |
| 250 | ±71 | 726.36 | +71 | 0.4914 |
| 300 | ±57 | 781.25 | +57 | 0.5627 |

(c)

| Width (W) | Optimized angle (α, °) | Optimized wavelength (λ, nm) | Circular dichroism (CD) | |
|---|---|---|---|---|
| 24 | ±63 | 599.63 | +63 | 0.1878 |
| 28 | ±55 | 621.30 | +55 | 0.2635 |
| 32 | ±45 | 650.05 | +45 | 0.2762 |
| 40 | ±71 | 726.36 | +71 | 0.4914 |

Fig. 3 (Color online) (a) Dependence of CD on the wavelength (λ) between 450 nm and 800 nm for 100 nm unit cell with α=77°, 150 nm unit cell with α=58°, 200 nm unit cell with α=45°, 250 nm unit cell with α=71°, and 300 nm unit cell with α=57°. The maximum CDs are 0.1345 at U=100 with λ=545.45 nm, 0.1976 at U=150 with λ=598.80 nm, 0.2750 at U=200 with λ=651.47 nm, 0.4914 at U=250 with λ=726.36 nm, and 0.5627 at U=300 with λ=781.25 nm, respectively. Highly oscillated curves are removed in U=250 and 300. (b) 100 nm, 150 nm, 200 nm, 250 nm and 300 nm unit cells coupled with optimised angle, wavelength and CD. (c) Optimised angle, wavelength and CD compared between 24, 28, 32, and 40 nm width (W) in the 250 nm unit cell size

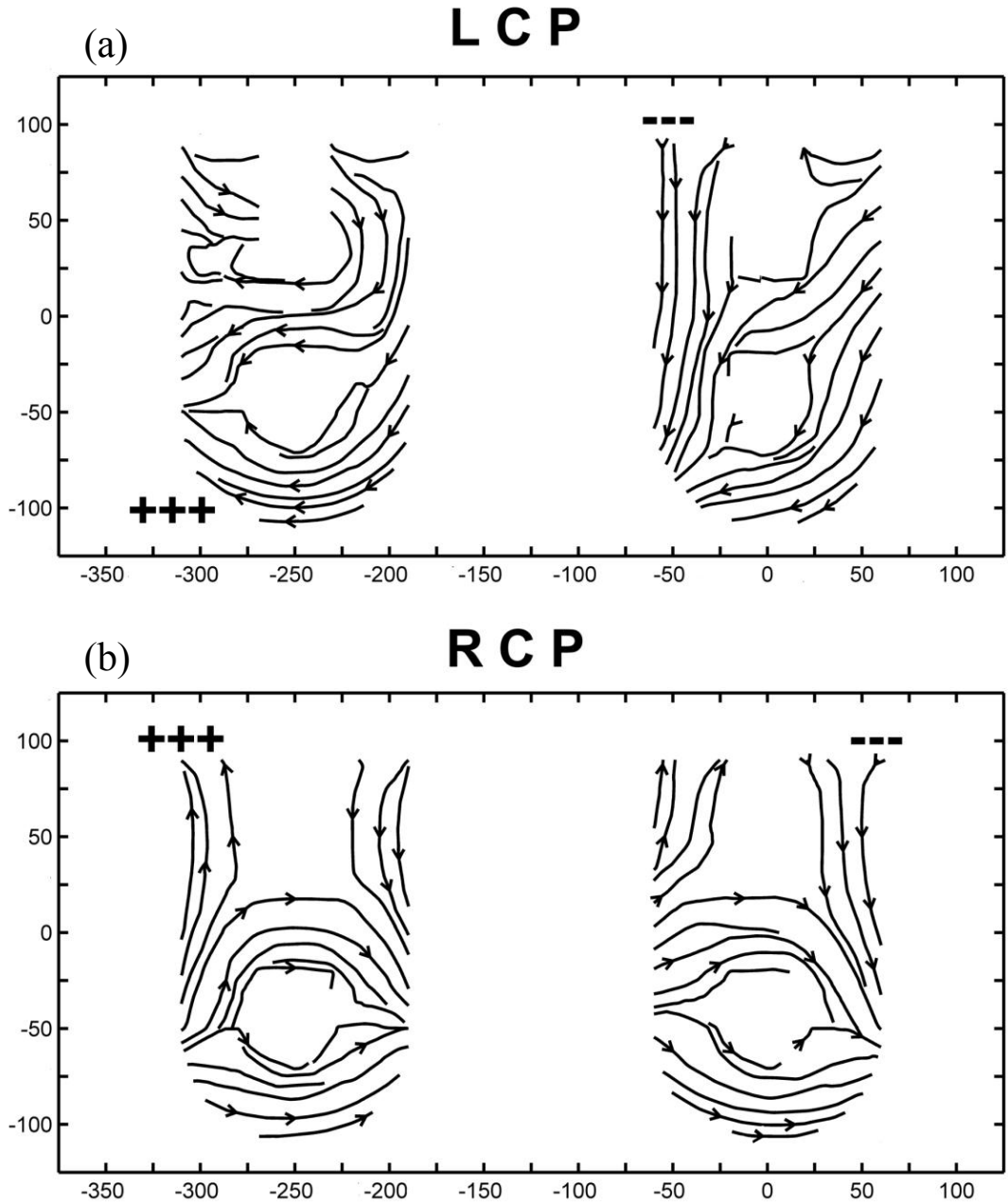

Fig. 4 (Color online) Current density field with streamline and vector direction in the 250 nm unit cell with 1000 nm height of borosilicate glass substrate, T=20 nm thickness, and W=32 nm width of silver unit cell at α=71° of incident beam and phase 0° with λ=726.36 nm between (a) LCP and (b) RCP.

clear optical activity. CD does not appear at 0 ° oblique incident beam for the reason that the projection of electric and magnetic moment is symmetrically the same between LCP and RCP *[14]*. The ±50 ° and ±80 ° intersects 0 ° plots are intersected at 700 nm wavelength but ±20 ° plot intersects 0 ° at 730 nm wavelength. All plots intersect again at about 1100 nm. It might cause optical activity with particular oblique incident beams. Each oblique incident beam generates different projection rotation of electric and magnetic diploes while k vector penetrates into the extrinsic 3D chirality.

A comparison of maximum CDs at 100 nm, 150 nm, 250 nm, and 300 nm unit cells with same pattern is shown in Fig. 3. When the unit cell size is decreased, the maximum CD is reduced and the peak wavelength is shifted to smaller wavelength. It may cause different electric and magnetic moment on the surface of the pattern. The bigger unit cell and wider structure has greater electric and magnetic energy because the optical activity can have more active resulting in the projection rotation of electric and magnetic diploes. Furthermore, the position and power of magnetic and electric dipoles may be relocated in ECMs. The 300 nm unit cell has the biggest CD but it is not in the visible spectrum so the 250 nm unit cell has the biggest CD in visible spectrum. Maximum CDs and these of angles are presented in Table 1. The 100 nm, 150 nm, 200 nm, 250nm, and 300 nm unit cells has the optimized angles at $\alpha= \pm77°, \pm58°, \pm45°, \pm71°$, and $\pm57°$, respectively. Optimized angles are independent of unit cell sizes because the electric and magnetic dipoles may be located in different positions and energy flow of optical activity may be changed on the surface of the pattern. The 250 nm unit cell with different width of the pattern is demonstrated in maximum CDs, that of an angle and wavelength as shown in Table 2. The maximum CD is 0.1878, 0.2635, 0.2762, and 0.4914 at W = 24, 28, 32, and 40, respectively. The smaller width can shrink total surface size of silver unit cells so the area of dipoles may be diminished. It may reduce the energy of surface plasmon on ECMs, and optical activity may not generate in smaller width strongly. The properties of CD could be lost in too big of width because the structure of ECMs is broken. A comparison of LCP and RCP at 250 nm unit cell with 71 ° tilted incident beam at phase 0 ° in current density field is shown in Fig. 4. The two unit cells were monitored simultaneously in Fig 4. (a) and (b). In the vertical incident beam, optical activity is not appeared between LCP and RCP so the LCP state of each unit cell is symmetrically opposite to the RCP. In 71 ° tilted incident beam, distribution of positive and negative charges is different between LCP and RCP at phase 0 °. Different energy distributions were also observed in the phase-averaged plots for LCP and RCP. Thus, physical mechanism of CD may be directly linked with different interactions between neighboring unit cells in LCP and RCP.

In conclusion, our ECM design can have strong CD with critical interactions between positive and negative charges on the surface of the patterns in visible spectrum. The maximum CD is generated by the mutual effect of tilted incident beams and ECM. The unit cell size has an impact on the CD and that of wavelength. Optimized angles do not depend on unit cell size and width of the pattern. Our ECM design can open up bio-molecular detecting systems and vibration sensors at visible spectrums.